\documentclass[12pt]{iopart}

\usepackage{hyperref}
\usepackage{graphicx}
\usepackage{soul}
\begin{document}

\title[]{Pitfalls in the teaching of elementary particle physics}

\author{Oliver Passon, Thomas Z\"ugge and Johannes Grebe-Ellis}

\address{Bergische Universit\"at Wuppertal, School of mathematics and natural sciences, Gau{\ss}str. 20, 42119 Wuppertal, Germany}
\ead{passon@uni-wuppertal.de}
\vspace{10pt}
\begin{indented}
\item[]October 2018
\end{indented}

\begin{abstract} Elementary particle physics is gradually implemented into science curricula at high school level. However, common presentations on educational, semi-technical or popular level contain or support severe misconceptions. We discuss in particular the notion of ``particle'', the interaction between them and the use of Feynman diagrams. In many cases the true novelty of particle physics (i.e. quantum field theory) is largely ignored.  We also suggest reasons for this widespread distortions of particle physics in popular accounts. 
\end{abstract}

%
%
%
%
%

\section{Introduction}

There are various arguments for including more recent science into physics curricula on high school level. Given the  coverage of e.g. particle physics, astrophysics or cosmology in the general media, the student's interest might be captured already and the  attractiveness of physics could be enhanced. 
Further more,  modern physics underpins many technologies  which deeply affect our daily lives. Another important point is to present physics not as ``finished'' but as an  on-going enterprise \cite[p. 88]{swinbank92}. This clearly needs the inclusion of some   cutting edge science into the curricula.  

Presumably for some of these reasons elementary particles and their interactions now form part of some syllabi on secondary school level. For example in the UK  particle physics has been a mandatory part of A level physics specifications since 2008 \cite{aqa,gourlay2017}. Already since the early 1990s it has been an optional module \cite{swinbank92}. In some German states it is a compulsory part (see e.g. \cite{nrw} {for the curriculum of  the most populous state, North Rhine-Westphalia}).  These syllabi include e.g. the composition of matter according to the Standard Model of particle physics and the concept of ``exchange particles''.

All this has given rise to many publications in the field of physics education which aim at presenting the Standard Model of particle physics on a level suitable to high school students (and their teachers, one might add). A rather general overview is given e.g. in \cite{farmelo92,ryder92,jones2002,hoekzema2005,daniel2006,vandenberg2006,organtini2011,hobson2011a,hobson2011b,johansson2013a,johansson2013b,woithe2017,kobel2017}.  Especially the use of Feynman diagrams as educational tool has attracted some interest. This specific aspect has been discussed e.g. in \cite{lambourne92,allday1997,dunne2001,pascolini2002,kontokostas2013}. Some of these articles address also the needs of teachers  on introductory undergraduate level. In addition there is a special category of papers which suggests to introduce particle physics even earlier, i.e. in primary school \cite{pavlidou2016} or to sixth graders  \cite{wiener2017}. In fact, we could have included the host of popular or semi-technical books on particle physics into our review since also they perform an educational function. However,  according to our anecdotal evidence this would not have altered our findings significantly. 

Note, that a significant fraction of these works do not originate from a genuine physics education research (PER) context. For example they rarely address alternative frameworks, conceptual change or relate their suggested curriculum to other areas taught. In addition, there is almost no literature on evidence-based PER in the area of particle physics. 
{Surprisingly, the specific educational value of particle physics is not addressed in the existing literature either. To fill this gap is the scope of one of our future projects} \cite{zuegge20xx}.

For the most part the authors of the papers we have investigated are particle physicists (mainly doing experimental work) who share their expertise with secondary school teachers, teacher trainers and those involved with courses on introductory undergraduate level. However, this may provide an important first step to the reconstruction of the subject. 

It turns out that beside of many differences in detail most of this presentations share a common core, that is, something like a  received view of ``simplified particle physics'' has emerged. According to this common narrative the elementary particles of the Standard Model, called quarks and leptons,  are the ``fundamental building blocks'' of  matter and the composition of hadrons and atoms out of these constituents is discussed. Most of what we know about particle physics originates from scattering experiments, i.e. the interaction between them and (finally) the detector material. These interactions are described by the exchange of other elementary particles; the so called gauge-bosons or exchange-particles (photons, gluons, W and Z particles). A common tool to visualize the interactions are so-called Feynman diagrams. In many presentations they  are introduced as representations of actual physical processes. Sometimes the Higgs mechanism (or rather  ``Englert-Brout-Higgs-mechanism'') is mentioned as well. It is needed to explain the masses of the elementary particles and the discovery of the corresponding Higgs particle at CERN in 2012 was the most recent triumph of this theoretical framework.  

Art Hobson is one of the few critics of this narrative which he calls the ``laundry list'' approach to teach the subject \cite[p. 12]{hobson2011a}. We will comment on his own approach in Sec.~\ref{par-sugg}. Also van den Berg and Hoekzema  focus on more conceptual issues ``instead of studying a multitude of particles and reactions'' \cite[p. 48]{vandenberg2006}. {The recent educational material of the German ``Netzwerk Teilchenwelt" explicitly avoids such a particle based approach} \cite{kobel2017} likewise. 

However, while for didactic reasons this presentation might be criticized, there are apparently no technical objections which can be made with regard to this sketch of the Standard Model, particularly given the need for brevity and dispense with mathematics. The expert will notice the use of many keywords and this way of talking about particle physics is indeed very close to the common jargon used at research facilities like CERN or SLAC. 

While this way of speaking might be unambiguous to the working physicist who has additional background knowledge and to whom this talk is embedded into a scientific practice, we will argue that to the layperson this narrative will almost  inevitable lead to severe misconceptions \cite[p. 48]{sms2002}. At the same time this account is omitting the true novelties of particle physics compared to non-relativistic quantum mechanics  or even classical physics. Novelties that in our opinion justify the inclusion into a high school science curriculum in the first place. 

However, in order to substantiate this claim we need to spell out what we take to be the ``true novelties'' of particle physics. We will do this along side with a more detailed discussion of the common presentation of this material in the papers cited above. In Sec.~\ref{particle} we take issue with the particle notion. Sec.~\ref{interaction} will deal with the use of Feynman diagrams and virtual particles to describe the interaction between the elementary particles. 
We close with some concluding remarks in Sec.~\ref{sum}. There we also take up the question why this widespread distortions of particle physics in educational, semi-technical or popular accounts could emerge.

\section{What are elementary particles?\label{particle}}
To all appearance ``particle physics'' deals with ``particles'', thus it seems reasonable at first to make this term more precise. We have surveyed the papers cited in the introduction and found a variety of strategies to introduce this key concept. A critical assessment of these suggestions will be made in Sec.~\ref{par-ass}.

\subsection{Suggestions from the educational literature\label{par-sugg}}
Woithe et al. \cite[p. 3]{woithe2017} state that the standard model is a quantum field theory and that the ``excitations of the field are identified as particles". However, they suggest not to speak of ``field excitations'' but simply of  ``elementary particle'' when discussing ``basic principles of particle physics qualitatively in high school". Also Allday in \cite[p. 327]{allday1997} introduces ``field excitations'' -- here, however, they are only used to characterize the gauge bosons (or ``exchange particles'', as Allday prefers to call them because this term is allegedly less intimidating). Allday explains  the interaction via boson exchange as a ``formation'' of these particles out of the underlying field and the photon as a ``disturbance'' of the field (p. 320).  Also Ryder  introduces ``field quanta'' \cite[p. 66]{ryder92} only to account for the gauge bosons. However, all this leads us into the subject of interactions which will be discussed in Sec.~\ref{interaction}. 

Farmelo \cite{farmelo92} is one out of many who speaks just of ``particles''  without any explanatory note. Johansson and Watkins introduce elementary particles as the ``smallest particles'' \cite[p. 105]{johansson2013b}. The notion of ``field excitation'' is mentioned only briefly in connection with the Higgs mechanism. Here the reader is rather abruptly told that the Higgs particle  is an ``excitation of the Higgs field'' (p. 112). In \cite[p. 97]{johansson2013a} Erik Johansson introduces the widespread metaphor of the particle collider as a gigantic microscope which allows to resolve structures in the range of the de Broglie wavelength $\lambda=h/p$ (which Planck's constant $h$ and $p$ the momentum of the beam particles). However, the attentive reader will be surprised to learn that ``according to de Broglie, the particle can be represented by a wave'' (p. 97). A further comment on these ``waves'' or the meaning of ``to represent a particle by a wave'' is missing and the paper continues to talk just about ``particles''. 

The microscope metaphor is also introduced by Jones \cite[p. 224]{jones2002}. Here, however, the nature of these ``mysterious'' (p. 224) waves, their probability interpretation  and the so-called ``wave-particle duality'' are briefly discussed. 
 Jones comes to the baffling conclusion:
\begin{quote}
 ``[...] although we do not understand these quantum waves as we would like, we will assume that, whatever properties other waves share, the quantum waves have them too.'' 
 \cite[p. 225]{jones2002}
 \end{quote}
In what follows Jones exploits the uncertainty principle and directs the discussion to cover the ``exchange model of forces'' and ``virtual particles''. Again, the discussion of these topics will be postponed until  Sec.~\ref{interaction}.  
 
Pascolini and Pietroni acknowledge that the notion of ``particle'' can not simply be taken over from classical physics. They write:
\begin{quote}
 ``Metaphors taken from the macroscopic world can convey only a very superficial account of what really goes on in the microscopic world, and they offer no access to the subtleties of genuine quantum behaviour.'' \cite[p. 324]{pascolini2002}
\end{quote} 
 These
authors try to sidestep this problem by introducing Feynman diagrams as ``accurate metaphors'' (p. 325). We will critically comment on this use of Feynman diagrams in Sec.~\ref{interaction}.

We have mentioned already the slogan ``elementary particles are excitations of the quantum field''. E.g. Woithe et al. \cite{woithe2017} introduce it briefly but suggest that it would be too technical to be used on high school level. As cited above, also Allday \cite{allday1997}, Ryder \cite{ryder92} and Johansson and Watkins \cite{johansson2013b} introduce this notion half-heartedly (at least for some ``particles''). Note, that if taken seriously, this slogan has important implications, namely that the ``particles'' loose their status as fundamental entities. To view them as field-excitations gives priority to the ``quantum field''. Exactly this point is emphasized by  Michael Daniel \cite[p. 120]{daniel2006}, when he writes:
\begin{quote}
 ``Quantization of a classical field is the formal procedure that turns a classical field into an operator capable of creating particles from vacuum. The quantization of the familiar electromagnetic field gives rise to particle-like quanta, the photons. [...] In this way we build up a theory, known as quantum field theory, in which the fundamental entities are the quantum
fields and not their quanta.'' \cite[p. 120]{daniel2006} 
\end{quote}
Note, that Daniel's wording is slightly different: he actually does not speak about ``field-excitation'' but emphasizes the even more technical aspect of ``particle creation''. We have more to say about this in Sec.~\ref{par-ass} and \ref{brief}. 

Perhaps the most distinguished representative of this ``all-fields view'' \cite[p. 211]{hobson2013} is Art Hobson. In a series of publications he has exploited this position already for non-relativistic quantum mechanics \cite{hobson2005,hobson2007,hobson2011,hobson2012}. In Ref.~\cite{hobson2013} with the programmatic title ``There are no particles, there are only fields'' this discussion is extended to quantum physics in general (i.e. including relativistic QT and quantum field theories). There, Hobson has collected compelling evidence against a coherent particle interpretation of quantum physics, ranging from QFT vacuum effects, non-locality and the ``no-go'' theorems by Hegerfeldt and Malament which point to the impossibility to localize objects in relativistic quantum physics. He concludes:
\begin{quote}
``There are overwhelming grounds to conclude that all the fundamental constituents of quantum physics are fields rather than particles. Rigorous analysis shows that, even under a broad definition of ``particle'', particles are inconsistent with the combined principles of relativity and quantum physics [...].'' \cite[p. 221]{hobson2013}
\end{quote}  
In \cite{hobson2011a,hobson2011b} Hobson addresses  the teaching of particle physics in particular. He argues against what he calls the ``laundry list'' approach to teach the subject, i.e. giving the mere list of ``elementary particles'' and its properties \cite[p. 12]{hobson2011a}. His own approach tries to convey what he takes the key concepts. This includes e.g. the existence of anti-particles as implied by the time-reversal symmetry of special relativity (p. 13f). With respect to the particle notion he certainly sticks to the slogan ``fields are all there is''.   While we agree with the destructive part of the argument (i.e. the shortcomings of the naive particle concept), the positive conclusion Hobson draws from it (``fields are all there is'') will be critically examined and questioned in Sec.~\ref{par-ass}. {As noted above, also the recent educational material of the ``Netzwerk Teilchenwelt'' is not based on particles but puts the main emphasis on ``charges'' and interactions. This points to a significant change in focus over the years, given  that the German ``Teilchenwelt'' means ``particle world''. They argue that the ordering scheme of the standard model can only be  appreciated  if especially the charge concept (i.e. weak- and color-charge) has been introduced. Otherwise, they suspect that the student could only memorize the different particle names} \cite[p. 87]{kobel2017}. 

\subsection{Critical assessment\label{par-ass}}
After this synopsis of the literature we try to place the suggestions in context. It is to be expected  that  simply using the common term ``particles'' without further explanation runs the risk to invoke the misconception  of  (very) small but ordinary pieces of matter. {There is a vast body of literature, mainly in chemistry education,  on student's problems to understand  the particle model. Here, ``particle'' is even understood classically, i.e. as used for example in the kinetic theory of gases.  Investigated have been applications like the expansion of a heated gas and  this research demonstrates clearly that already this classical particle model is rarely understood properly and that many students (on high school and college level) hold alternative frameworks; see e.g.} \cite{novick,treagust}.

This problem is particularly clear when elementary particles are just introduced as the ``smallest particles'' (Johansson and Watkins \cite[p. 105]{johansson2013b}). The accelerator-as-microscope metaphor is dangerous for the same reason, since it suggests that these particles can be actually ``seen''  -- what else should a microscope be good for. All this supports the everyday idea of ``particles''. The microscope-metaphor is also problematic on a technical level. Note, that the resolution-limit in microscopy is, broadly speaking, due to diffraction effects at the aperture. Nothing similar applies for the detection of elementary particles.

To equate ``elementary particles'' tacitly with this everyday idea leads into all kind of problems.  As noted already by Barlow in 1992 \cite{barlow92} the static quark model of the hadrons appears simple and intuitive at first, e.g. if the proton is describe as an $(uud)$ state. However, already the neutral pion ($\pi^0$) as a ``mixture'' of $u\bar{u}$ and $d\bar{d}$ provokes problems if concepts of quantum mechanics can not be assumed ({or have been introduced only cursory}). Barlow sees only two alternatives:
\begin{quote}
``[...] (a) avoiding the subject and hoping nobody asks awkward questions or (b) telling them that when they go to university the truth will be revealed to them there. Neither of
these is very satisfactory.'' \cite[p. 93]{barlow92}
\end{quote}

Thus, Barlow expresses his concern about whether or not it would be possible to present particle physics not too simplistic, given that  (at least in some countries) quantum mechanics  is not covered until undergraduate level.
The same problem is acknowledged by Allday:
\begin{quote}
 ``Increasingly these days teachers are having to present ideas that really need quantum mechanics without the luxury of being able to teach quantum mechanics as a subject. I am thinking especially of the rise of particle physics.'' \cite[p. 327]{allday1997}
 \end{quote}  
Against this background the probability-waves introduced by Jones \cite[p. 224]{jones2002} seem to be a step in the right direction. However, the theoretical framework of the Standard Model is not non-relativistic quantum mechanics but quantum field theory. There, in general, such a probability interpretation does not exists. This point will be explained in more detail below. 

Leaving for the moment  the problem of an appropriate educational reconstruction aside one might think that ``particles'' as ``excitations of the quantum field'' as introduced e.g. by Daniel  \cite{daniel2006} or Hobson \cite{hobson2013} capture the notion most accurately.  However, also this  common {\em fa\c{c}on de parler} turns out to be problematic. While in all other areas of physics a ``field" is understood as a function which assigns to each space-time point a scalar, vector or tensor valued quantity, the solutions of the ``quantum field equations" are operator valued. It follows that they have to be applied to a state vector to result into anything which can be interpreted physically. The quantum field as such (i.e. the operator valued quantity) cannot have any ``excitation".  Thus, strangely enough, one may argue whether ``quantum field theory" is a ``field theory" at all. In Sec.~\ref{brief} we will provide some more technical details on quantum field theory. 

Michael Daniel in \cite[p. 120]{daniel2006} is one of the few authors we could find in the educational literature who emphasizes exactly the operator-valuedness of the quantum field. However, he jumps to the conclusion that the quantum field is the primary object of this theory -- apparently ignoring the problems indicated above (compare our quote from Daniel in Sec.~\ref{par-sugg}). 

The same criticism applies to Hobson. His ``all-fields view'' \cite{hobson2013} has been commented on by Sciamanda \cite{sciamanda2013}, de Bianchi \cite{debianchi2013} and Nauenberg \cite{nauenberg2013}. While  de Bianchi and Nauenberg take mainly issue with aspects of non-relativistic quantum theory or the definition of localizability respectively, Sciamanda addresses the operator-valuedness of the quantum field. His argument is similar to the one mentioned above. 

Sciamanda, in his reply to Hobson,  concludes that QFT ``does little or nothing to identify a model for the ontological reality [...] of the entity being described'' (p. 645). In closing he remarks:
\begin{quote}
``Perhaps there is no useful conceptual model to describe ultimate reality in human terms -- and perhaps there is no need for one.''
\end{quote}
We believe that Hobson's reply \cite{hobson2013b} misses the point when he accuses Sciamanda to align with those who claim that reality ``fades from existence'' at small distances. Sciamanda's point is more modest when he questions that we can describe reality in ``human terms'', as quoted above. 
The case against the field interpretation of QFT has been further strengthened  by Baker in \cite{baker2009}.

In discussing the view of particles as ``field excitations'' we came across  a number of technical terms  from quantum field theory (``operator-valued fields'', ``creation operators'' or ``Fock-space''). In order to assess these suggestions of the educational literature more thoroughly we need to introduce some concepts of quantum field theory slightly more systematically. In particular we would like to indicate    how ``particles'' are introduced into this formalism and what properties they share with particles as understood in classical physics.

\subsection{A brief look at ``particles'' in quantum field theory\label{brief}}
A common strategy to turn a classical field theory  (like e.g. electrodynamics) into a   quantum field theory (quantum electrodynamics (QED) in this case) is the so-called  ``canonical quantization".\footnote{Note, that the term ``canonical'' refers to the role played by ``canonical variables'' in a Hamiltonian formulation of the classical theory that forms the starting point of the construction.This approach to field quantization has been criticized for its lack of mathematical rigor. Among other reasons this inspired the development of algebraic quantum field theory (AQFT) \cite{aqft}. In \cite{bertozzi2013} this highly technical approach is used to draw lessons for the teaching of particle physics. However, in its present form it  appears very idiosyncratic to us.} Here one promotes the classical fields to be operators and demands certain commutation relations between these operator-fields and their ``conjugate momenta" (defined as the partial derivative of the Lagrangian with respect to the time derivative of the field). 

If one expands the solution of the free quantum-field equation into a Fourier series the coefficients of the expansion have to turn into operators likewise  (for technical details on the quantization of the electromagnetic field see e.g.~\cite{gerry2005} or any other text on theoretical particle physics, quantum field theory or quantum optics).  Given the commutation-constraints mentioned above these ``coefficients" turn out to satisfy the same algebraic relations as the raising and lowering operators in the quantum mechanical treatment of the harmonic oscillator. In the context of quantum field theory these operators are called creation ($a^{\dagger}(p)$) and annihilation ($a(p)$) operators.

It is this appearance of the harmonic oscillator which motivates the talk about ``excitations", given that an oscillator may be excited indeed. However, the analogy is purely formal since the coefficients of the previously mentioned expansion happen to obey the same algebraic relations. One can not possibly view the ``quantum field" as a set of harmonic oscillators. As we will just show,  the corresponding states lack almost every properties which one would associate with ``particles". 

As in the case of the quantum mechanical harmonic oscillator, the combination $N(p)=a^{\dagger} (p)a(p)$ is a hermitian operator (called ``occupation number operator") which possesses a discrete spectrum of eigenvalues: 0, 1, 2, 3, $\cdots$. The corresponding eigenstates are accordingly called the ``number states". They form a basis of the state-space of the interaction-free theory, called the Fock-space of the system.\footnote{The Fock-space is the (direct) sum of $N$-particle Hilbert spaces for $N=0, 1, 2, 3, \cdots$.} For example the state with lowest eigenvalue $|0\rangle$ is called the ``vacuum state''. 

Applying the creation operator on such a Fock-state raises its occupation number by one.  For example   acting on the vacuum state leads to a state with occupation number ``1'' and momentum $p$; formally: $a^{\dagger}(p)|0\rangle\sim|1\rangle$. Thus, in QED, this is an ``one-photon state''.\footnote{The QFT description of ``matter fields'' proceeds, {\em cum grano salis}, in the same way. Here, the starting point is the solution of the relativistic Dirac equation. However, instead of requiring certain ``commutation relations'' one needs ``anti-commutation relations'' to meet the requirement of Fermi-Dirac statistics. The corresponding quantum fields describe all elementary fermions. This leaves only open how the novel interactions (weak and strong) enter the formalism. Expressed technically, QED possesses a so-called gauge symmetry. If one requires a similar symmetry (with different ``gauge groups'' though) one arrives at a description of the weak and the strong interaction. What has been the ``photon'' for quantum electrodynamic is now called the ``W and Z bosons'' (for the weak interaction) and  the ``gluons'' (for the strong interaction). However, this ``local gauge invariance'' requires the corresponding quanta to be massless. The actual masses need an additional mechanism -- due to Higgs et al.  What this sketchy account neglects is that actually the Standard Model introduces a unified framework for the electromagnetic and the weak interaction. For this and any other detail one needs to consult a textbook on particle physics or quantum field theory, e.g. \cite{peskin}.} 

Note, that it is only the discreteness of the spectrum of the number-operator which justifies the talk about ``particles", i.e. discrete entities. It is this discreteness which implies  discrete values for energy and momentum (for eigenstates of the number operator) likewise  -- and this fact is apparently too seductive to avoid talking about ``particles''. However, any other connotation of this term does not apply. Already in ordinary quantum mechanics the notion of a particle trajectory has to be rejected. However, the notion of a ``particle position" remains applicable in QM since a position operator exists. In contrast, the discrete ``photon-states" of QED cannot be localized at all since no position operator for them can be constructed \cite{newtonwigner49}. As noted by many, the ``fuzzy ball" picture of the photon fails \cite{scully72,strnad86,kidd89,jones91}.\footnote{We note for completeness that a position observable for the photon may be defined if one moves to generalized observables which are not self-adjoint anymore; so-called ``positive operator valued measures" (POVM) \cite{kraus}. Meanwhile, the study of quantum theory with such generalized observables has turned into an industry. However, Halvorson and Clifton \cite{halvorson2002} have shown the limitations for this POVM approach to localization for relativistic quantum theory. } 
 It is very unfortunate, that according to the common teaching tradition a particle-like photon is included already into the quantum mechanics curriculum in order to explain the photo-electric effect. In the light of the later developments this revolutionary proposal by Einstein in 1905 was too classical still. Einstein suggested a localized and distinguishable ``quantum of light" which should not be confused with the current photon concept of QED; compare also the discussion in \cite{passon2017}.

 In textbooks on theoretical particle physics these issues do rarely take center stage, since their goal is to provide different skills. However,  e.g. Peskin and Schroeder remark at this point:
\begin{quote}
``
By a {\em particle} we do not mean something that must be localized in space; $\hat{a}^{\dagger}_k$ creates particles in momentum eigenstates.'' \cite[p. 22]{peskin}
\end{quote}

This turns out to be an exciting lesson indeed. While ``classical particles" (say, within the kinetic theory of gases) have number, position, a trajectory and can be distinguished, the ``particle states" of quantum mechanics possess discrete number, position (although in general not sharply defined), no trajectory and are indistinguishable. Moving one step further into quantum field theory leaves one essentially with only one feature left: discrete number. However, also this feature is in general not sharply defined, since an arbitrary state will be a superposition of different number-eigenstates. Note, that this counter-intuitive feature allows to describe the typical high energy physics reactions in which particles are created and destroyed.\footnote{However, there is an additional blow to the particle interpretation, even in this weakened form. Doreen Fraser has shown, that in {\em interacting systems} no Fock-space representation can be given \cite{fraser2009}. Consequently, only for the asymptotic scattering limit (i.e. before and after the scattering interactions) an interpretation of the states in term of discrete entities (some times called ``quanta'') can be given.} As mentioned  above, Hobson rejects the notion of ``particles'' as fundamental  constituents for similar reasons. 

Obviously, this  sketch of quantum field theory is not intended to become part of a physics curriculum on high school level. We just made this technical detour to illustrate how insufficient any naive talk about ``elementary particles" is. If one  ignores the lessons of both, quantum mechanics and quantum field theory on any level of presentation, severe misconceptions will follow. Introducing such a concept of ``particle" may  lead the students to think about them as small pieces of {\em ordinary matter} -- a finding, as indicated above, which has been reported frequently when testing the student concepts with regard to the particle model \cite{novick,treagust}. 



\section{Interactions, Feynman diagrams and virtual particles \label{interaction}}
Most of what we know about and the elementary particles comes from scattering experiments and  the predictions of the theory regard the probability for observing specific final states, given a specifically prepared initial state. This information is encoded in the corresponding scattering cross-section. To calculate  a cross-section within the Standard Model is very demanding. As a rule it can not be derived analytically and most predictions are obtained within perturbation theory, i.e. the solution is approximated by an expansion in powers of the coupling constant.  

In the 1940s Richard Feynman  introduced a graphical tool (named after him ``Feynman diagrams'') to facilitate the calculations considerably. Within Feynman diagrams  each particle state is graphically displayed  by a line. The points where two or more lines meet are called ``vertices'' and contribute (among other things) a ``coupling constant'' (i.e. some generalized charge) to the calculation. Further more inner and outer lines need to be distinguished.
 
For obvious reasons these  diagrams fascinate many people in the educational context. They carry the promise that complex mathematics can be avoided without distorting the underlying physics. Only few authors of the papers we have investigated could resist the temptation to introduce Feynman diagrams as a tool to visualize and to illustrate particle interactions. In using this approach one is lead naturally to introduce ``(virtual) exchange particles'' as the mediators of the interaction. 
Among the question which have to be addressed are:
\begin{enumerate}
\item How close is the relation between  Feynman diagrams and the underlying physical processes? 
\item How can an interaction (attractive or repulsive) be mediated by some ``particle exchange'' at all? 
\item What is the status  of the infamous ``virtual particles''? 
\end{enumerate}
The reply to these questions also  depends on the  level of sophistication with which the particle concept is introduced, i.e. relates to the discussion of Sec.~\ref{particle}. 
Again, we will first review the suggestions from the literature before we turn to a critical assessment in Sec.~\ref{int-ass}.

\subsection{Suggestions from the educational literature\label{int-sugg}}
\subsubsection{On the relation between Feynman diagrams and the underlying physical process}

As noted already in Sec.~\ref{par-sugg}  Pascolini and Pietroni  acknowledge the problem of intuitive pictures and introduce Feynman diagrams  as ``accurate metaphors'' \cite[p. 325]{pascolini2002} (this expression appears close to self-contradictory). They have build an actual  mechanical model of the components of a Feynman digram (straight sticks for electrons, cylinders with appropriate ports for the vertices etc.) and emphasize the manipulative aspect of the resulting ``game'' as a key feature to the success of this project which was aiming to  convey concepts like ``antimatter, conservation laws, particle creation and destruction, real and virtual particles'' to high school students. Their view on the relation between Feynman diagrams and the underlying physical process is stated clearly:
\begin{quote}
``Each diagram -- constructed according to well-defined rules -- represents a possible physical process and, making it so valuable to physicists, it can be unambiguously translated into a mathematical expression, giving the probability for that process.''\cite[p. 325]{pascolini2002}
\end{quote}
This is a typical example of what we will call the ``literal reading'' of Feynman diagrams. Also Johansson and Watkins illustrate the interactions between elementary particles with the help of Feynman diagrams. Like all the other authors they emphasize the mathematical complexity, which is concealed behind them:
\begin{quote}
``The mathematics behind a diagram is very detailed. Each external line represents a real particle, and the Feynman rules define how the propagation of the particle is described mathematically.'' \cite[p. 107]{johansson2013b}
\end{quote}
However, their wording (``propagation of particles'') is very vivid and may be taken to view Feynman diagrams as illustrations of the actual particle path in space-time. In fact, some of the papers we investigated equip Feynman diagrams with a space-axis; e.g Jones calls them ``two-dimensional spacetime graphs'' \cite[p. 229]{jones2002} and Kontokostas and Kalkanis \cite{kontokostas2013} speak of vertices as points were ``world lines'' meet. 


In Ref.~\cite{woithe2017} the introduction to the standard model is framed into the explanation of each term of the Lagrange density. This approach is apparently very distant from the students perspective, however, the authors argue that this mathematical formula is printed on T-shirts and coffee mugs at the CERN gift shop and high school students may come across it. The central strategy in the discussion of Woithe et al. \cite{woithe2017} is to translate each term of the Lagrangian into the respective Feynman diagram (or rather fundamental vertices). This formal tool is introduced with some care. The authors note   that each diagram is just a pictorial representation of a mathematical expression which describes a particle interaction. They point to the danger that they might falsely be taken to represent a  motion and emphasize that only a time-axis can be specified. However, they also claim that a Feynman diagram 
\begin{quote}
``[...]  is a useful tool to visualise and understand what is happening in a certain interaction without the need for mathematics." \cite[p. 3]{woithe2017}
\end{quote}
 When discussing the beta-decay or the vector-boson fusion process Woithe et al.  clearly indicate that also here each Feynman diagram is taken to represent an actual physical process. 

A much more careful presentation is given by Jonathan Allday. He emphasizes that in general no single diagram represents a physical process and states pointedly: ``[...] no single diagram is happening -- they are all involved'' \cite[p. 331]{allday1997}. He suggests to compare the situation with the superposition of different motions in classical mechanics and we will explore this example in some more detail in Sec.~\ref{int-ass}.  Another example for a more modest claim about the relation between Feynman diagrams and the underlying physical process is this remark by Lambourne: 
\begin{quote}
``Almost no one can resist the temptation of using Feynman's diagrammatic expansion to justify talk of an incoming electron `emitting' a virtual (i.e. internal to a diagram) photon that is then `absorbed' by an outgoing electron. While giving in to such temptation is perfectly understandable, I hope you can now see that there is much more to these diagrams than popular magazine articles usually make clear.'' \cite[p. 74]{lambourne92}
\end{quote}
However,  he calls this temptation ``understandable'' and remarks that there is ``much more'' to these diagrams. This leaves open how justified and coherent this literal reading of Feynman diagrams actually is. Apparently many authors feel uncomfortable to read too much into the   diagrams. A particularly telling example is provided by Jones. At first he introduces Feynman diagrams as two-dimensional space time graphs and indicates that even the angle at which a line is drawn would contain information  \cite[p. 229]{jones2002}. In closing the corresponding subsection he offers the following vague restriction:
\begin{quote}
``One should not try to take this picture too literally and try to give a running commentary of the evolution of the intermediate state. The picture just tells us that we have imagined the two-stage process I gave you as an example.''  \cite[p. 229]{jones2002}
\end{quote}
Also Organtini comments on the interpretation of Feynman diagrams with a  puzzling and ambiguous remark: 
\begin{quote}
``The Feynman diagram in figure 1 can be thought of as the visual representation of a process in which an electron coming from the bottom left, emits a particle represented by $A$ and, consequently, modifies its direction towards the top left, to conserve momentum. It is important to understand that this is not necessarily what happens at the microscopic level. The reality is described by the equation of motion, not by the Feynman diagram. The latter is just another, funny way to write $\alpha\overline{\psi}A\psi$'' \cite[p. 546]{organtini2011}
\end{quote}
The first part of this statement takes the Feynman diagrams apparently to describe actual particle trajectories. We do not even need to reproduce the figure he refers to, since his description is complete. In any event he seems to favor a literal reading.  In closing, any space-time description or representational function of Feynman diagrams is denied in favor for the interaction-term encoded in the mathematical formalism. This puzzle is partly resolved when Organtini notes a few lines further down that due to the successful predictions derived from perturbation theory:
\begin{quote}
``[...] we are free to interpret Feynman diagrams as what happens at the microscopic level and we are led to the conclusion that interactions can be represented by the exchange of mediators, even if we cannot see them.''\cite[p. 547]{organtini2011}
\end{quote}
It is certainly common to ground the ``reality'' of theoretical terms (like ``quarks''  or ``black holes'') on the predictive success of the related theories. In fact, we think that we have good reasons to believe in the reality of ``quarks'', ``leptons'', ``gauge bosons'' and the like. However, with respect to the detailed mechanism of particle-interactions this claim is based on a simple misconception; see Sec.~\ref{int-ass}. 

\subsubsection{Interactions mediated by ``exchange particles''}

All this  leads us finally to a discussion of  the  issue of interactions being mediated by their respective exchange particles, as mentioned in the above quote and present in most of the papers we have investigated. Obviously, this question is closely related to the issue  whether and how Feynman diagrams relate to the underlying physical process since  the very notion of ``exchange particle'' is essentially the meaning attached to the internal lines of these diagrams.\footnote{Note, however, that all types of particles may be represented by internal lines. E.g. in describing Compton scattering at lowest order the photons are external   and the internal line represents a ``virtual'' electron. On the notion of ``virtual particles'' see below.} 

The level of sophistication with which this notion can be introduced and explained depends largely on the issue discussed in Sec.~\ref{par-sugg}. On a naive or everyday notion of ``particle'' 
the ``particle-exchange'' turns into a mechanical model with very limited scope.  E. g. Farmelo introduces interactions via exchange particle with the well known analogy of either throwing a ball (repulsive force) or a boomerang (attractive force) \cite[p. 100]{farmelo92}. Jones defends this way to illustrate the mechanism:
\begin{quote}
``Some people do not like this analogy. It does, however, have two features that make it useful: (i) It is thought-provoking and memorable, even as a piece of classical physics. (ii) It involves spin, and although this takes us beyond the scope of this introductory paper, it is now known that the force-carrying particles -- the photon, the intermediate vector bosons Z and W, and the gluon -- all have an intrinsic angular momentum or spin.'' \cite[p. 230]{jones2002}
\end{quote}

Allday is one of  the few authors who address the critcal issues for the learner (not only) on high school level. One of them is: ``How can the exchange of a particle give rise to an attractive force as well as a repulsion?''\cite[p. 327]{allday1997}.  Allday comments on the boomerang-analogy (which is also introduced by \cite{kobel2017}) sarcastically:
\begin{quote}  
``The most absurd way of explaining this that I have seen is to accuse the exchange particle of being like a boomerang that curves away from the emitter in the opposite direction to what one would expect. It then whips round behind the target and knocks it towards the emitter.''\cite[p. 330]{allday1997}
 \end{quote}
He  points out that this presentation misses the true nature of the Feynman diagrams completely. In fact,  $e^-e^-$ scattering (``repulsive'' interaction) and $e^+e^-$ scattering (``attractive'' interaction) are represented by the {\em same} Feynman diagram, or, to be more precise:  one of the two tree-level diagrams (``the $t$-channel exchange of one photon'') for these processes is identical. Only if one considers all relevant diagrams (at given order) the difference between  attraction and repulsion can be explained. But this, Allday continues, can not be pictured with any simplified particle model but needs the consideration of the field aspect. He takes this to be the main problem with Feynman diagrams:
\begin{quote}  
``Feynman diagrams hide the fact that there is already `contact' between the two particles before the exchange takes place. The Feynman diagram does not show the field that underlies the process. Understanding this is vital to feeling comfortable about the whole Feynman approach.''\cite[p. 330]{allday1997}
 \end{quote}
 Clearly, this is related to his view on ``particles'' as discussed in Sec.~\ref{par-sugg}. There we noted that the ``exchange particles'' are introduced by Allday as field-excitations, the interaction being described as a formation of gauge bosons out of the underlying field. 

\subsubsection{The notion of ``virtual particles''} 

Finally we need to turn to the infamous ``virtual particles''. As noted by many authors the exchange particles fall into this category. Technically, ``virtual particles'' are represented by the internal lines of Feynman diagrams, which do not have to obey the energy-momentum relation:
\begin{eqnarray}
E^2=p^2c^2+m^2c^4. \label{rem}
\end{eqnarray}
These ``particles'' are said to be ``off mass shell'' or simply ``off shell''. This notion  adds even more complexity to the already obscure presentation of interactions in particle physics. Presumably only few people would follow  Pascolini and Pietroni in their bold remark:
\begin{quote}
``The concept of virtual particles is self-explanatory in the language of Feynman diagrams; it is sufficient to draw the user's attention to the difference between those lines that have a free
end (real particles) and those where both ends terminate on a vertex (virtual ones).''\cite[p. 326]{pascolini2002}
\end{quote}
This is a rather operational definition of the term and gives neither any deeper insight nor relates the notion with other concepts already acquired by the ``user''.  In order to provide exactly such a connection several authors resort to the energy-time uncertainty relation\cite{jones2002,hobson2011b,johansson2013a,johansson2013b}: 
\begin{eqnarray}
\Delta E \cdot \Delta t \ge \frac{\hbar}{2} \label{t-e}
\end{eqnarray}
According to Jones this  relation implies:
\begin{quote}
``[...] that we can take seriously the possibility of the existence of energy non-conserving processes -- provided the amount by which energy is not conserved, $E_{\mathrm{violation}}$, exists for a time less than $h/4\pi E_{\mathrm{violation}}$. This idea will then form the basis of a discussion of the Exchange Model of Forces.'' \cite[p. 226]{jones2002}
 \end{quote}
In connection with the massive gauge bosons of the weak interaction this energy-violating effects are invoked to explain the exchange even if the incoming particles are much lighter (and less energetic) \cite{jones2002,hobson2011b}. Further more the short range of these interactions should be explained by the short time span $\Delta t$ available. Again, we will comment on this widespread explanation in Sec.~\ref{int-ass}. Among the few critics is Allday who remarks with tongue in cheek:
\begin{quote}
``Whenever a physicist turns to the uncertainty principle to explain anything I start to get deeply worried. A fudge is about to be perpetrated.''\cite[p. 329]{allday1997} 
\end{quote}
 An other rare example for a more modest claim with respect to ``virtual particles'' is also this quote by Daniel: 
\begin{quote}
``  `Virtual particles' are not observable quantities, and in fact the concept of a virtual particle is useful only to the extent that it allows us to develop an intuitive understanding of interactions between particles. [...] 
Within the framework of quantum field theory, however, it is more accurate to think of particle interactions as a direct consequence of quantum field interactions giving rise to elementary vertex interactions.'' \cite[p. 126]{daniel2006}
\end{quote}

 
\subsection{Critical assessment\label{int-ass}}
As indicated briefly at the beginning of this section 
 Feynman diagrams depict terms in the expansion of elements of the so-called scattering matrix $\langle f|S|i\rangle=S_{fi}$ in powers of the coupling constant. Here, $i$ and $f$ denote the initial and final state of the scattering respectively. The matrix $S$ encodes the interaction terms of the underlying theory and is defined as \cite[p. 81]{lahiri2000}:
 \begin{eqnarray}
S = {\cal{T}} \left  [  \exp \left( -i \int_{-\infty}^{+\infty} d^4x {\cal{H}}_I \right ) \right ]. 
\end{eqnarray}
Here, ${\cal{T}}$ denotes the time-ordering operator and ${\cal{H}}_I$ the interaction part of the Hamiltonian.  Expressed technically, the sum of all diagrams to a given order $n$ represents a term $S_{fi}^{(n)}$  in the expansion
\begin{eqnarray}
S_{fi}\approx S_{fi}^{(0)}+g S_{fi}^{(1)}+\cdots +g^N S_{fi}^{(N)}.
\end{eqnarray}
 Here,  $N$ denotes the order of the approximation, i.e. the highest power of the vertex factor $g$ entering. This factor $g$ is related to the coupling constant by $g \propto \sqrt{\alpha}$ (e.g. in QED $\alpha$ is the fine structure constant, i.e. $g$ is proportional to the electron charge. If the spin is taken into account the vertex factor includes an additional term.). 

Now, the probability of the process to occur is proportional to the square of the scattering matrix element:
\begin{eqnarray}
P_{i\rightarrow f} \propto |S_{fi}|^2\approx \left | S_{fi}^{(0)}+g S_{fi}^{(1)}+\cdots +g^N S_{fi}^{(N)} \right |^2. \label{square} 
\end{eqnarray}

\subsubsection{Feynman diagrams and their representational function}
The innocent looking square in Equ.~\ref{square} (or actually the square of the modulus of this complex number) is key to clear up some confusions about the interpretation of Feynman diagrams. 
To interpret a {\em single term} of this sum physically ignores the effects which underlie already the double-slit experiment in quantum mechanics. There, in a very similar way, the {\em square} of the sum of the contributions of the different slits yields the observable interference pattern and the question whether the electron ``actually" went through slit 1 or 2 cannot be asked, let alone answered. In other words: the observed interference   is related to the mixed term, i.e. the observable effect can not be related to a single term alone \cite{redhead88,weingard}. {In the teaching of quantum mechanics  it is usually a central objective that no space-time picture of the process between preparation and measurement can be given. It is curious to note that on a literal reading of Feynman diagrams some authors apparently accept the risk  to fall  below this level.}

All this is closely related to the point Allday has made in \cite{allday1997} (compare our remark in Sec.~\ref{int-sugg}). He compares the different contributions of the Feynman diagrams for a given process with the superposition of different motions in classical mechanics. In his example he considers a pendulum in which the thread is replaced by a spring. It can be set in motion such that the pendulum motion superimposes with the oscillation of the spring. Allday  compares each motion in isolation with a {\em single} Feynman diagram -- the actual  motion is a complicated mixture of both. In this way he clarifies the already quoted claim: ``[...] no single diagram is happening -- they are all involved'' \cite[p. 331]{allday1997}. At this point the reader might object that while no {\em single} diagram should be taken to represent the underlying process, the imagined {\em sum} would do so -- as in the classical motion example of Allday. However, what is missing in this oscillator example is the fact that in the case of the Feynman diagrams the observable phenomenon is related to the {\em square} of the sum. This gives rise to interference terms which are missing in Allday's mechanical analogy. So the case against a literal reading of Feynman diagrams (individually or summed) can be made even stronger. 



Note in addition, that the cross-section finally deduced from the $S$-matrix gives just the probability to observe the transition from certain initial into final states. Any individual interaction between specific objects can not possibly be expressed by it. Presumably in the same vein Gell-Mann remarks: 
\begin{quote}
``In QED, as in other quantum field theories, we can use the little pictures invented by my colleague Richard Feynman, which are supposed to give the illusion of understanding what is going on in quantum field theory." \cite[p. 170]{gell-mann}
\end{quote}
We mentioned Organtini's claim that the empirical success of the perturbative predictions warrants the literal reading of Feynman diagrams. As shown above such a reading is actually at odds with the application of this tool (e.g. the ``square-after-sum'' rule). Thus, it is exactly the success of the method which should guard us against such an interpretation. 

Other, more technical arguments, can be given against the literal reading of Feynman diagrams. The notable property  of  Feynman diagrams is, that they allow a perturbative expansion which satisfies Lorentz invariance at each stage of the calculations. This, however, needs all terms (to given order) to be taken into account. A related issue is gauge symmetry. The Standard Model interactions can be derived from the requirement of invariance under certain gauge-symmetry operations. Any physical prediction needs to be ``gauge independent''. In general individual diagrams do not fulfill this requirement and only the sum is always gauge invariant \cite[p. 512]{ricardo}.  Strangely enough, on a literal reading of individual Feynman diagrams one actually tries to explain the working of a relativistic gauge-theory while violating both, relativity and gauge invariance.   

\begin{figure}
	\centering
	\includegraphics[width=0.95\textwidth]{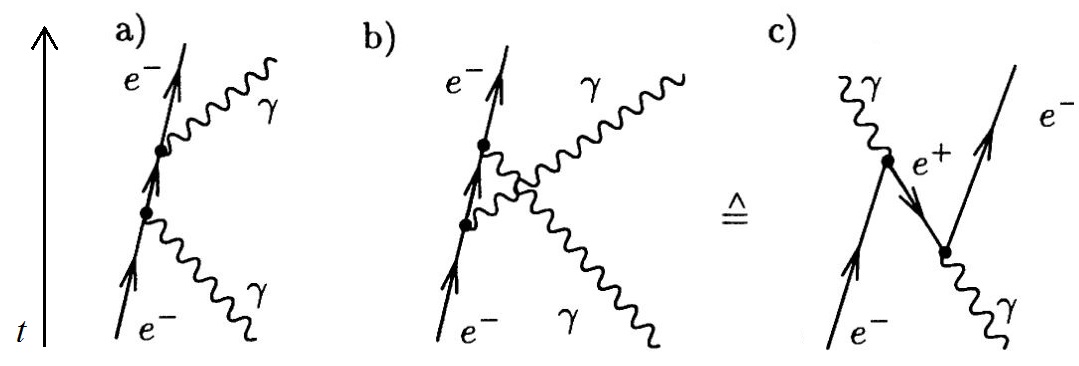}
	\caption{Feynman diagrams for the  Compton scattering in lowest order. In diagram a) a photon is first absorbed and subsequently emitted. The diagrams b) and c) differ only by time-ordering and are ``topological equivalent'', i.e. it is sufficient to consider just one of them.
	}
	\label{FD}
\end{figure}
There is an additional argument against the representational function of Feynman diagrams which derives from the so-called ``topological equivalence''. Feynman diagrams which can be continuously deformed into each other while leaving the in- and out coming states unchanged are called ``topologically equivalent'' \cite[p. 68]{mandlshaw}. They represent the same amplitude and only one member of this equivalence class has to be considered. Let us look at the example of Feynman diagrams depicted in Fig.~\ref{FD}.  
They show the different contributions to the Compton scattering ($e^-\gamma\rightarrow e^-\gamma$) in lowest order perturbation theory. Fig. a) seems to capture the ``intuitive sequence of events'': first one photon is absorbed and subsequently one photon is emitted. However, already in lowest order an other term needs to be considered as well. This is the amplitude depicted in b) where photon emission and absorption reverse their time-order. This appears already counter-intuitive. However, an other -- topological equivalent -- diagram is shown in c). Here, the incoming photon splits into $e^+e^-$ while the incoming electron annihilates with the positron to give rise to the final state photon. The stories which can be told to describe the diagrams b) and c) are quite different, but given their equivalence only one needs to be considered. They represent only different ways to draw the {\em same} diagram.  This aspect of time-ordering has also been noted by Allday \cite[p. 328]{allday1997} and Daniel \cite[p. 125]{daniel2006}, however they do not draw any conclusion from it.

A further obstacle against any literal reading of Feynman diagrams  -- especially in terms of a space-time story --  lies in the fact that they do not have a space-axis. Strictly speaking even a time-axis does not exist -- the usually indicated time-arrow serves only the function to distinguish between incoming and outgoing states. A quick glance at the Feynman rules in any QFT textbook shows, that the length of the lines, the angle at which they meet or the position of the vertices do not carry any information about the corresponding mathematical expression.

Thus, there are major problems to sustain  the view that Feynman diagrams represent any underlying physical process. This, in particular, affects the notion of ``particle exchange'' as mediating interactions. St\"ockler suggests that this should be viewed as a metaphor only \cite[p. 242]{friebe2015}. For similar reasons James Brown concludes that Feynman diagrams ``do not picture any physical processes at all. Instead, they represent probabilities (actually, probability amplitudes)'' \cite[p. 265]{brown1996}. 
He compares them with Venn diagrams which also provide the visualization of an abstract relation (``being part of''), without claiming that the little circles visualize the corresponding system in any other respect.\footnote{A more detailed discussion on the interpretation of Feynman diagrams in the context of the recent debate within the philosophy of physics can be found in \cite{passon-int}.}

\subsubsection{The status of virtual particles}
Let us finally turn to the   ``virtual particles'', i.e. states which do not meet the relativistic energy-momentum condition \ref{rem} and which are represented by internal lines in the Feynman diagrams. As noted in Sec.~\ref{int-sugg} many authors invoke the energy-time uncertainty to make this plausible and explicitly claim that this relation implies a violation of energy conservation.  This argument is disturbing for several reasons, as noted already by Bunge in 1970  \cite{bunge70}.  For one thing we do not know of any sound argument for the statement that in quantum mechanics energy conservation is violated. {Famously, Bohr, Kramers and Slaters suggested in 1924  that energy and momentum may not be conserved in each single light-matter interaction but only statistically. This suggestion was soon disproved experimentally by Bothe and Geiger} \cite{jammer}.

{As noted in the last section, some authors claim that the energy-time uncertainty relation implies the short range of the weak interaction, i.e.  an }{\em upper} bound for the time span $\Delta t$, if the violation of energy conservation $\Delta E$ is large (say, in the order of the $Z$ or $W$ mass). This, however,  would need a $\le$-sign instead of the $\ge$-sign in Equ.~\ref{t-e}. {One way to save the argument is to quote the energy-time uncertainty in the form} $\Delta E\Delta t\approx \hbar$. {However, successful applications of the energy-time uncertainty relation deal e.g.  with the energy-spread of states (say, the natural line width) and their lifetime, i.e., have nothing to do with any alleged energy non-conservation.} 

But even if the laws of quantum mechanics were to allow a violation of energy conservation, the issue in relativistic quantum field theory should be the violation/conservation of energy-momentum anyway. Now, according to the Feynman rules the constraint of energy-momentum {\em conservation} is met at each vertex. Thus, it is just the other way around: The internal lines representing ``off-shell'' states is a {\em consequence} of exactly this energy-momentum {\em conservation} -- rather than being implied by any violation of this principle. We can only speculate about the reasons for the flawed  standard account. Perhaps these authors think in terms of the so-called old-fashioned (or time-ordered) perturbation theory (compare e.g. \cite[Chap. 4]{schwartz2014}). In this formalism the time-ordering of the vertices needs to be considered, i.e. one distinguishes between retarded and advanced propagators. At each vertex 3-momentum is conserved but the energy is not. However, this kind of violation of the energy conservation principle in intermediate steps of the calculation is not related to ``virtual states'' since in this formalism all states are on-shell all the time \cite[p. 52]{schwartz2014}. Another reason for the widespread notion of energy violating intermediate states might be that loop diagrams need the integration over the loop momentum. However, apparently this argument is uncritically applied to tree diagrams as well. 

All this certainly leaves open how to think about this off-shell states instead. For example one might view ``virtual photons'' as some exotic states with $m\neq  0$ (although this would violate gauge invariance).  In fact, in calculating the scattering cross-section one needs to sum over the time-like and longitudinal  polarization states, i.e. those states which  ``real photons'' do not have \cite[p. 503]{ricardo}. 

However, our previous discussion has raised doubts on a literal reading of Feynman diagrams anyway. If they are taken as calculational device only, one should not attach any deeper meaning to this  intermediate (and unobservable) states. This view is further supported by the following fact. The recursive method to solve field equations with the help of Green's functions can also be applied for classical field equations. Also here the expansion can be depicted by ``Feynman diagrams'' with external and internal lines -- thus there is nothing  particularly ``quantum'' about them. In fact, this observation make it seem plausible that ``virtual particles'' are just artifacts of a specific solving technique (perturbation theory).  If e.g. the  numerical solution technique of lattice gauge theory is applied no ``virtual particles'' emerge.

\section{Summary and conclusion\label{sum}}

We have investigated a number of educational sources which are intended to help teachers on high school or introductory undergraduate level to introduce the basic concepts of elementary particle physics. While we found a broad agreement in the general outline, there are confusing many incoherences or even inconsistencies on important question. {Scarcely if at all discussed questions are}: Is it possible to adopt a ``particle interpretation'' of quantum field theory? Should the ``particles'' be rather viewed as  ``excitations of the quantum field'' or is some even more radical break with this notion advisable? How do Feynman diagrams relate to the underlying physical process, in particular:  How exactly is the slogan ``interactions are mediated by the exchange of (virtual) particles'' to be understood? Admittedly, these are complex issues which are debated  within physics and philosophy of physics. But to simply bypass them does no service to teachers and students likewise. 

{Most of the presentations we have surveyed introduce a simplistic notion of particles and interaction via particle-exchange}. However, if particle physics is just presented as a set of novel and exotic words for elementary building blocks which replace the atoms of chemistry or protons and neutrons of nuclear physics it fails to serve any educational purpose. In that case it simply reiterates a strategy which has been witnessed by the student several times already (a similar argument is given by Kobel et al. \cite[p. 87]{kobel2017} to justify their interaction-based approach). We fully agree with Art Hobson who has dubbed this procedure ``the laundry list approach''. With tongue in cheek he remarks that this conventional approach ``is a great way to make a fascinating topic meaningless'' \cite[p. 12]{hobson2011a}.

One might argue that this complex topic needs suitable reduction in order to be taught on high school level. Clearly, any area of physics is simplified accordingly and to demand scholarly acceptable presentations is off the mark. However,   to teach this topic is only justified if at least {\em some} of the novel aspects are introduced and if no misconceptions are supported or even produced. A suggested curriculum which tries to meet these requirements will be presented in \cite{future}. 

This paper has traced  a widespread narrative (not only) in the educational literature. Even Murray Gell-Mann, whom we quoted previously for his critical remark concerning Feynman diagrams,  continues his Wolfson lecture \cite{gell-mann} along the same lines. Also the popular introductory textbook by Povh et al. \cite{povh} introduces Feynman diagrams as pictorial representations of physical processes and even equips them with a space axis (p. 50). So one might wonder why distinguished scholars and experts in the field contribute to the trivialization of this area when it comes to educational, popular or semi-technical accounts. We can only speculate as to the reasons, but some of the following suggestions seem plausible. 

For one thing, there is an ``operational" meaning of particles as objects which can be   approximately localized within particle detectors and which are bundles of properties (like mass and charges). No ontological commitment is needed when dealing with these objects as   practicing physicist \cite[p. 220ff]{brigitte}. But as suggestive as the pictures of particle tracks in bubble chambers or rather reconstructed tracks from the data of modern particle detectors might be, one should avoid the pitfall to view them as the effect of a local cause. Already on the level of non-relativistic quantum mechanics this conclusion can not be supported \cite[p. 4ff]{heisenberg30}.  

We have discussed some reasons why it is problematic to picture light as being composed of photons or quarks as constituents of hadrons -- at least in any conventional understanding of the part-whole relation. However, this narrative is still strong, presumably since  the quark model originated exactly from a solution to the ``particle zoo'' problem.  That is, the starting point was a discovery which allows for the reduction of this complexity by a simple ordering-scheme. If this historical path is followed one is naturally lead to the conclusion, that the elements of this (static) ordering scheme are material constituents.  

In addition, high energy physics prepares matter in a state which makes the intuitive particle description apparently suitable. The high energies of the colliding beams at, e.g.,  LHC allows for the description of asymptotically free states in the scattering process, i.e. even quarks in the final states can be viewed as free states. The subsequent process of ``hadronisation" escapes the perturbative description and can only be accounted for with the help of so-called fragmentation models which are of a heuristic nature. 

A further reason might be that some of these problematic accounts are produced by outreach groups of research facilities. Their goal is not only educational and the activity aims also to justify the research towards the  funding agencies and eventually the tax payer. Perhaps in this context a less abstract presentation appears more beneficial.

Also textbooks of particle physics on university level introduce the ``particle-jargon", this time supplemented  by the corresponding mathematics. The goal here is rather to develop technical skills, like the calculation of cross-sections. Thus, they need not comment on the issue whether e.g. a single Feynman diagram refers to any physically realized process.  But all this misleading jargon and metaphors are embedded into a scientific practice which usually protects the scientists to be led astray. For example no matter how they {\em talk} about Feynman diagrams -- in the end they will {\em apply} the correct  Feynman rules. Now, if the scientific practice is stripped off and only the jargon taught to students, there is less protection against these  misconceptions.

Finally and presumably the single most important reason for this widespread distortions is, that the question how the objects of quantum field theory are embedded into space-time is unresolved yet. Thus, there is no intuitive space-time story to be told in order to illustrate the results of particle physics and the same applies,  albeit in alleviated terms, for quantum mechanics already. That is, some of the apparently  indispensable tools in physics teaching, namely intuitive pictures and simple models, can not easily be applied here. Apparently Feynman diagrams are introduced by some as a substitute for such models and we have discussed at length why this strategy is misguided. 

However, models might be misleading in other areas of physics likewise. Stephen Toulmin once remarked with respect to the model of light rays that the sentence ``light travels in straight lines" only appears to be a statement about some newly discovered property of light. In fact, nobody initially believed that light would travel in curved lines instead. This statement actually assigns a new meaning to the term ``light" and even extends the notion of ``to travel'' \cite[p. 20]{toulmin}. {Apparently the same is true for the statement ``elementary particles are the building blocks of matter". Also this remark gives a new meaning to the term ``particle" and the concept of ``being made of''.} 

We believe that to resolve this meanings and to question the ordinary narrative of part-whole relations  at least to some extend is the actual goal one should strive for if particle physics is included into science curricula. To suggest such an approach will be part of a future project \cite{future}.

\section*{Acknowledgment}
This project is supported by the ``Qualit\"atsoffensive Lehrerbildung'' a joint initiative of the Federal Government and the federal states which aims to improve the quality of teacher training. The programme is funded by the Federal Ministry of Education and Research. The authors are responsible for the content of this publication (FKZ: 01JA1507).


\end{document}